\documentclass[twocolumn]{aastex631}
\pdfoutput=1

\def\bfa{Bayes factor}
\def\bfs{Bayes factors}
\def\blue{\ac{UV} and optically blue}

\def\distances{\ac{lumdist} = \SI{40}{Mpc}, \SI{100}{Mpc} and \SI{160}{Mpc}}
\def\D{\textit{Dorado}}

\def\default{`default'}
\def\bright{`lower early opacity'}
\def\data{photometric data}

\def\gbo{ground-based optical}
\def\gw{GW170817}
\def\gfo{AT2017gfo}
\def\grb{GRB170817A}
\def\kn{kilonova}
\def\knmodel{kilonova model}
\def\knmodels{kilonova models}

\def\red{optically red and \ac{NIR}}
\def\sbu{satellite-based \ac{UV}}
\def\sh{shock interaction}
\def\shmodel{shock interaction  model}

\usepackage{savesym}
\savesymbol{tablenum}
\usepackage[range-units = single]{siunitx}
\DeclareSIUnit\parsec{pc}
\restoresymbol{SIX}{tablenum}
\usepackage{amsmath}
\usepackage{bm}
\usepackage{graphicx}
\usepackage{booktabs}
\usepackage{xcolor}
\usepackage[nolist]{acronym}

\begin{acronym}
\acro{1D}{one dimensional}
\acro{2D}{two dimensional}
\acro{3D}{three dimensional}
\acro{BH}{black hole}
\acro{BHNS}{black hole neutron star}
\acro{BBH}{binary black hole}
\acro{BNS}{binary neutron star}
\acro{CBC}{compact object coalescence}
\acro{CCD}{charge coupled device}
\acro{DETC}{Dorado exposure time calculator}
\acro{DOS}{Dorado observation scheduler}
\acro{EFE}{Einstein field equations}
\acro{effwav}[$\lambda_\mathrm{eff}$]{effective wavelength}
\acro{EM}{electromagnetic}
\acro{EoS}{equation of state}
\acro{ETC}{exposure time calculator}
\acro{FoV}{field of view}
\acrodefplural{FoV}{fields of view}
\acro{GCN}{$\gamma$-ray Coordinates Network}
\acro{GRMHD}{general-relativistic magneto-hydrodynamic}
\acro{GRB}{$\gamma$-ray burst}
\acro{GW}{gravitational wave}
\acro{GWB}{gravitational wave background}
\acro{HLV}{Advanced LIGO and Virgo}
\acro{HLVK}{Advanced LIGO, Virgo and KAGRA}
\acro{HPDI}{highest posterior density interval}
\acro{IR}{infra-red}
\acro{ISM}{interstellar medium}
\acro{LIGO}{Laser Interferometer Gravitational-Wave Observatory}
\acro{LCO}{Las Cumbres Observatory}
\acro{lumdist}[$d_L$]{luminosity distance}
\acrodefplural{lumdist}[$d_L$]{luminosity distances}
\acro{LVC}{LIGO-Virgo collaboration}
\acro{MMA}{multi-messenger astronomy}
\acro{NIR}{near infra-red}
\acro{NS}{neutron star}
\acro{PPD}{posterior probability distribution}
\acrodefplural{PPD}{posterior probability distributions}
\acro{S/N}{signal-to-noise ratio}
\acro{SDSS}{Sloan Digital Sky Survey}
\acro{SED}{spectral energy density}
\acrodefplural{SED}{spectral energy densities}
\acro{sGRB}{short $\gamma$-ray burst}
\acro{SED}{spectral energy distribution}
\acro{ToO}{target of opportunity}
\acrodefplural{ToO}{targets of opportunity}
\acro{UV}{ultra-violet}
\acro{UVO}{\ac{UV} and optical}
\acro{UVOIR}{\ac{UV}, optical and \ac{IR}}
\end{acronym}
\usepackage[noabbrev]{cleveref}
\makeatletter
\usepackage{etoolbox}
\patchcmd\H@refstepcounter{\protected@edef}{\protected@xdef}{}{}
\makeatother
\usepackage{subfigure}
\graphicspath{{./}{figures/}}

\received{XXXXXX}
\revised{XXXXXX}
\accepted{\today}

\shorttitle{}
\shortauthors{Dorsman et al.}

\begin{document}

\title{Prospects of Gravitational Wave Follow-up Through a Wide-field Ultra-violet Satellite: a \textit{Dorado} Case Study}

\newcommand{\GRAPPA}{\affiliation{GRAPPA, Anton Pannekoek Institute for Astronomy and Institute of High-Energy Physics, University of Amsterdam, Science Park 904, 1098 XH Amsterdam, The Netherlands}}
\newcommand{\Nikhef}{\affiliation{Nikhef, Science Park 105, 1098 XG Amsterdam, The Netherlands}}
\newcommand{\Goddard}{\affiliation{Astroparticle Physics Laboratory, NASA Goddard Space Flight Center, Mail Code 661, Greenbelt, MD 20771, USA}}
\newcommand{\Caltech}{\affiliation{Division of Physics, Mathematics, and Astronomy, California Institute of Technology, Pasadena, CA 91125, USA}}
\newcommand{\tokyo}{\affiliation{Research Center for the Early Universe, Graduate School of Science, University of Tokyo, Bunkyo-ku, Tokyo 113-0033, Japan}}
\newcommand{\Dirac}{\affiliation{DIRAC Institute, Department of Astronomy, University of Washington, 3910 15th Avenue NE, Seattle, WA 98195, USA}}
\newcommand{\Carnegie}{\affiliation{The Observatories of the Carnegie Institution for Science, 813 Santa Barbara St., Pasadena, CA 91101, USA}}
\newcommand{\Clemson}{\affiliation{Department of Physics and Astronomy, Clemson University, Clemson, SC 29634-0978}}

\correspondingauthor{Bas Dorsman}
\email{b.dorsman@uva.nl}

\author[0000-0002-9407-0733]{Bas Dorsman} \GRAPPA

\author[0000-0002-9397-786X]{Geert Raaijmakers} \GRAPPA

\author[0000-0003-1673-970X]{S. Bradley Cenko} \Goddard

\author[0000-0001-6573-7773]{Samaya Nissanke} \GRAPPA \Nikhef

\author[0000-0001-9898-5597]{Leo P. Singer} \Goddard

\author[0000-0002-5619-4938]{Mansi M. Kasliwal} \Caltech

\author[0000-0001-6806-0673]{Anthony L. Piro} \Carnegie

\author[0000-0001-8018-5348]{Eric C. Bellm} \Dirac

\author[0000-0002-8028-0991]{Dieter H. Hartmann} \Clemson

\author[0000-0001-5023-6933]{Kenta Hotokezaka} \tokyo
 
\author[0000-0002-0332-7899]{Kamilė Lukošiūtė} \GRAPPA

\begin{abstract}
The detection of gravitational waves from binary neuron star merger \gw{} and electromagnetic counterparts \grb{} and \gfo{} kick-started the field of gravitational wave multimessenger astronomy. The optically red to near infra-red emission (`red' component) of \gfo{} was readily explained as produced by the decay of newly created nuclei produced by rapid neutron capture (a kilonova). However, the ultra-violet to optically blue emission (`blue' component) that was dominant at early times ($\lesssim$\,1.5 days) received no consensus regarding its driving physics. Among many explanations, two leading contenders are kilonova radiation from a lanthanide-poor ejecta component or shock interaction (cocoon emission). In this work, we simulate \gfo{}-like light curves and perform a Bayesian analysis to study whether an ultra-violet satellite capable of rapid gravitational wave follow-up, could distinguish between physical processes driving the early `blue' component. We find that a \textit{Dorado}-like ultra-violet satellite, with a 50 deg$^2$ field of view and a limiting magnitude (AB) of 20.5 for a 10 minute exposure is able to distinguish radiation components up to at least 160 Mpc if data collection starts within 3.2 or 5.2 hours for two possible \gfo-like light curve scenarios. Additional sensitivity and additional filters may allow for a longer acceptable response time. We also study the degree to which parameters can be constrained with the obtained photometry. We find that, while ultra-violet data alone constrains parameters governing the outer ejecta properties, the combination of both ground-based optical and space-based ultra-violet data allows for tight constraints for all but one parameter of the kilonova model up to 160 Mpc. These results imply that an ultra-violet mission like Dorado would provide unique insights into the early evolution of the post-merger system and its driving emission physics. In addition, this study shows that ultra-violet plus optical multi-wavelength detections provide complementary constraints, jointly covering a broader range of early ejecta properties.
\end{abstract}


\section{Introduction} \label{sec:introduction}
The \ac{LVC} can now regularly detect and study \acp{GW} from the final moments of inspiraling compact object mergers \citep[][]{Abbott2020}. At least some compact binary mergers involving a \ac{NS} are expected to produce \ac{EM} counterparts, depending on mass-ratio and spins \citep[for a review, see e.g.][]{Nakar2020}. The discovery of a \ac{BNS} merger \gw{} \citep{Abbott2017c} and subsequent \ac{EM} counterparts, the \ac{GRB} \grb{}  \citep[][]{Goldstein2017, Savchenko2017} and \ac{UVOIR} counterpart \gfo{} \citep[e.g.,][]{Abbott2017a, Abbott2017b, Arcavi2017, Chornock2017, Coulter2017,Kasliwal2017,Lipunov2017,Shappee2017, Soares-Santos2017, Valenti2017}, kick-started the era of \ac{GW} multimessenger astronomy. \gfo{} exhibited a featureless thermal spectrum, peaking in the near-\ac{UV} at $\sim$\,1 day (`blue' component). After this time the blue component faded while the \red{} emission (`red' component) shifted further towards the \ac{IR} \citep[e.g.,][]{Arcavi2017, Cowperthwaite2017,  McCully2017, Nicholl2017, Shappee2017}.

It has long been predicted that the ejecta of \ac{BNS} or \ac{BHNS} mergers can host a thermal transient powered by the radioactive decay of heavy elements formed by rapid neutron captures (r-process), and these astrophysical transients have been named macronovae or kilonovae \citep[][]{Lattimer1974, Li1998, Kulkarni2005, Metzger2010}. While the red component of \gfo{} is consistent with predictions from several kilonova models \citep[][]{Kasen2013, Tanaka2013, Rosswog2017, Wollaeger2018} the blue component is hotter, more luminous, and faster rising than what is predicted by the aforementioned models \citep{Drout2017, Kasliwal2017}. One way to explain the full light curve evolution is by considering the presence of multiple distinct ejecta components \citep[e.g.,][]{Kasen2017, Tanaka2017, Villar2017}. In such a scenario, ejecta with a higher electron fraction ($Y_e \gtrsim$\,0.25) would produce material with a lower lanthanide fraction and subsequently brighter, bluer, day-long emission. Such an ejecta component with higher electron fraction could correspond to neutrino driven or magnetized winds from a long lived \ac{NS} central remnant \citep[e.g.,][]{Shibata2017, Metzger2018, Nedora2021} but in the case of \gfo{} such winds would have difficulties explaining the high velocities inferred for the early emission. Still, \cite{Waxman2018} found that a single ejecta component kilonova model, but with a uniform time dependent opacity ($\kappa \propto t^{\gamma}$), is consistent with the available data. Alternatively, other model ingredients may be driving the blue component such as a precursor powered by free neutron decay \citep[][]{Metzger2015, Gottlieb2020} or shock interaction, also known as cocoon emission \citep[][]{Kasliwal2017, Piro2018}.

One critical aspect that was missing in the observations of \gfo{} was \ac{UV} data in the first hours. \textit{Swift}/UVOT obtained their first exposure in the ultraviolet band ($\sim$\,200 -- 300 nm) at $\Delta t = 0.6$ days (15 hours) after the \ac{GW} trigger \citep{Evans2017} of \gw. Early time \ac{UV} data would provide the essential diagnostic to identify the main mechanism or combination of mechanisms that power the \blue{} radiation in the first few hours \citep{Arcavi2018}. One way to obtain such data would be by having a large \ac{FoV} \ac{UV} satellite on stand-by to rapidly follow up on \ac{GW} triggers to find the target in the \ac{GW} localization sky area while it is still bright in \ac{UV} in the first few hours. Two missions that have been proposed to fulfill this purpose are ULTRASAT \citep[][]{Sagiv2014} and UVEX \citep[][]{Kulkarni2021}. In \Cref{sec:discussion}, we discuss the applicability of our findings to these currently planned \ac{UV} missions. We also note that recently \cite{Chase2022} performed a kilonova detectability study for a wide range of wide-field instruments.

Performing early \ac{UV} follow-up of \ac{GW} events was also an objective of \textit{Dorado}, a mission concept that was submitted to the 2019 NASA Astrophysics Mission of Opportunity call for proposals. This study assumes \textit{Dorado}'s follow-up and observational capabilities when simulating \ac{UV} photometry. Summary mission specifications of \textit{Dorado}, ULTRASAT and UVEX are given in \Cref{tab:missionspecs} and we discuss \textit{Dorado} and the simulation of \data{} in detail in \Cref{sec:data}.

\begin{deluxetable}{cllll}
    \tablecaption{\label{tab:missionspecs}Mission specifications of mission concept \textit{Dorado} and proposed missions ULTRASAT and UVEX. From left to right, the table presents the 5$\sigma$ limiting magnitude (AB) for given exposure time, response time, \ac{FoV}, and orbit type. LEO, GEO, and HEO are low earth, geo-synchronous, and high earth orbit respectively.}
    \tablecolumns{5}
    \tablewidth{0pt}
    \tablehead{
       \textbf{Mission} & \textbf{5$\sigma$ (AB)} & \begin{tabular}{@{}l@{}}\textbf{Response} \\ \textbf{time}\end{tabular} & \textbf{$\Omega$ (deg$^2$)} & \textbf{Orbit}
    }
    \startdata
    \textit{Dorado} & \begin{tabular}{@{}l@{}}20.5 \\ (10 min. exp.)\end{tabular} & $\sim$\,30 min.   & 50  & LEO \\
    ULTRASAT      & \begin{tabular}{@{}l@{}}22.3 \\ (15 min. exp.)\end{tabular} & $\sim$\,30 min.   & 200 & GEO \\
    UVEX            & \begin{tabular}{@{}l@{}}25 \\ (15 min. exp.)\end{tabular}   & $\lesssim$\,3 hr. & 12  & HEO \\
    \enddata
\end{deluxetable}

In this study, we perform a Bayesian analysis to examine to what degree \sbu{} photometry could distinguish between two competing emission models for \gfo-like blue emission. The two models considered are r-process nucleosynthesis (\kn) and \sh{} (cocoon emission). Currently, these two options are prominent in the literature, and the data of \gfo{} allows for either a pure \sh{} or \kn{} scenario to explain the blue component. In addition, we study to what degree the data constrains parameters of either model. We also perform an analysis with \gbo{} and joint \ac{UVO} data to examine the added value of early \ac{UV} photometry. Finally, we study how delayed observation affects model selection and parameter estimation for \sbu{}.

This paper is structured as follows: \Cref{sec:radiationmodels} describes the kilonova and shock interaction models. \Cref{sec:methodology} describes the Bayesian theoretical framework and the simulation of light curves and \data. \Cref{sec:results} presents the results of the Bayesian analysis. \Cref{sec:discussion} presents our main findings, implications for planned \ac{UV} missions, discusses caveats for this work, and concludes.

\section{Radiation Models}\label{sec:radiationmodels}
This section provides a summary of the physics of the two radiation models and discusses the model parameters.

\subsection{Nucleosynthesis Powered Model}
The first radiation model is the \citet[][]{Hotokezaka2020} model for radiation powered by $\beta$-decay of radioactive elements produced through r-process nucleosynthesis (kilonova). This is a semi-analytic model, based on the Arnett model \citep[][]{Arnett1982}. The latter models the light curves for supernovae and as such includes the radioactive decay of $^{56}$Ni and $^{56}$Co. \cite{Hotokezaka2020} applies to kilonovae however, and solves the time evolution of $\beta$-decay chains for all elements up to a certain mass $A$ to get the radioactive power of each decay chain. For simulations performed here elements up to $A=209$ are included. Here, we do not include the heating due to $\alpha$-decay and fission because their contributions are rather small in the early times, where kilonovae are expected to be bright in UV. Thermalization is treated similar to analytic methods presented by \cite{Kasen2019, Waxman2019}, with injection energies of each decay product specified per decay chain. 

The model assumes isotropic geometry where radial density structure of the ejecta is modeled as a function of time $t$ and velocity $v$, given by

\begin{equation}
    \rho(t,v)=\rho_0(t)\left(\frac{v}{v_{\rm min}}\right)^{-n} \quad (v_{\rm min} \leq v \leq v_{\rm max}),
\end{equation}
where $\rho_0(t)$ ensures that the full $M_{\rm ej}$ is retrieved when the density $\rho(t,v)$ is integrated over the full velocity range [$v_{\rm min}, v_{\rm max}$]. Parameter $n$ defines how steeply the power-law density distribution falls off. The ejecta are modeled here as finely discretized mass shells, and a notable feature is that radiative transfer is improved by accounting for radiation escape from mass shells with different expansion velocities in the case where the diffusion time is long compared to the dynamical time. 

We also note that the model uses a concentric piece-wise opacity distribution. While the \cite{Hotokezaka2020} model allows up to two opacities, our adaptation expands the model to allow for an arbitrary number of $m$ zones. An array of $m+1$ velocities defines the borders of these expanding opacity zones, specifying each transition velocity. Nevertheless, in this work we use only two opacity zones, with inner opacity $\kappa_{\rm in}$ and outer opacity $\kappa_{\rm out}$. The velocities are $v_{\rm min}$, $v_{\rm transition}$ and $v_{\rm max}$. Note that bound-bound transitions of heavy elements dominate the kilonova opacities, which in reality vary with the wavelengths and ejecta conditions such as temperature and density \citep[e.g.,][]{Tanaka2020,Banerjee2020}. Here, however, we assume opacity to be constant with time and wavelengths for simplicity. An effective radius and temperature is calculated such that the model produces effective blackbody radiation. Our adaptation of the model is publicly available on GitHub\footnote{\url{https://github.com/Basdorsman/kilonova-heating-rate/}}.

\subsection{Shock Interaction Powered Model}
The second radiation model is an analytical \shmodel{} derived by \cite{Piro2018}. It follows analytical work presented in \cite{Nakar2017} and includes some added details to facilitate comparison with \gfo. In this model, a relativistic \ac{GRB} jet punches through ejecta material, depositing some of its energy to the ejecta, and thereby inflating it to create a shock heated cocoon. The light curve is powered by cooling emission of this shock-heated material. Such a model could also represent a shock driven by a short-lived magnetar \citep{Metzger2018}. Similar to the \knmodel, this model contains no angular dependence and emits effective blackbody radiation.

The shock model assumes an initial shock radius $R$ corresponding to the radius of the ejecta at the time when the \ac{GRB} first punches through. For \gw{}, $R$ is $\sim$\,$\num{e10}-\num{e11}$ \si{cm}, corresponding to a jet traveling at $\sim$\,$c$ taking about $\sim$\,1.7 seconds to emerge, as derived from the delay time between \gw{} and \grb{} \citep[][]{Abbott2017a, Goldstein2017, Savchenko2017}. The shocked envelope is assumed to have energy distributed with respect to velocity as

\begin{equation}
    \label{eq:dEdv}
    \frac{dE}{dv}\propto v^{-s},
\end{equation}
where it is assumed that $s>-1$. The parameter $s$ encodes ignorance on how exactly the jet deposits its energy in the ejecta. As in \cite{Piro2018}, we assume $s=3$ which is derived from \gfo. The luminosity $L$, effective temperature $T_\mathrm{eff}$ and effective radius $r_\mathrm{eff}$ are given by

\begin{align} \label{eq:Lshock}
    L &= \num{9.5e40} \kappa_{0.1}^{-3/5}M_{0.01}^{2/5} v_{0.1}^{8/5}R_{10} t_\mathrm{day}^{-4/5} \mathrm{\; erg\ s}^{-1},\\ \label{eq:Tshock}
    T_{\mathrm{eff}} &= \num{6.2e3} \kappa_{0.1}^{-7/30}M_{0.01}^{1/60}v_{0.1}^{1/15}R_{10}^{1/4} t_\mathrm{day}^{-8/15} \mathrm{\; K},\\ \label{eq:rshock}
    r_{\mathrm{eff}} &= \num{3e14} \kappa_{0.1}^{1/6}M_{0.01}^{1/6}v_{0.1}^{2/3}t_\mathrm{day}^{2/3} \mathrm{\; cm},
\end{align}
where $M_{0.01}$ is the shocked mass in units of $0.01 M_\odot$, $\kappa_{0.1}$ is the opacity in units of \SI{0.1}{cm^2/g}, $v_{0.1}$ is the minimum ejecta velocity in units of 0.1\,c, $R_{10}$ is the initial shock radius in units of \SI{e10}{cm}, and $t_\mathrm{day}$ denotes the days elapsed since merger.

\section{Methodology}\label{sec:methodology}
We perform a Bayesian analysis to quantify the degree to which various photometric data distinguishes the radiation models and allows for parameter estimation. This section elaborates on our methodology and is structured as follows: \Cref{sec:bayes} provides a short summary of the relevant theory of Bayesian model selection and parameter estimation. The Bayesian analysis takes as input observed data and, in this study, that data is simulated in two steps. Firstly, \Cref{sec:lightcurves} defines `fiducial' (i.e. \gfo{}-like) light curves computed with the models introduced in \Cref{sec:radiationmodels}.  Secondly, \Cref{sec:data} lays out how an observation of these light curves is simulated, producing simulated \data.

\subsection{Bayesian framework}\label{sec:bayes}
Bayesian model selection and parameter estimation \citep[for an in-depth review, see e.g.][]{Sivia2006} rely on some data $D$, and a model $M$ described by a set of parameters $\theta$. The posterior probability is given by

\begin{equation}\label{eq:posterior}
    P(\theta|D,M)=\frac{P(D|\theta,M)P(\theta|M)}{P(D|M)}=\frac{\mathcal{L}(\theta)\pi(\theta)}{\mathcal{Z}},
\end{equation}
where $\mathcal{L}(\theta) = P(D|\theta,M)$ is the likelihood function and $\pi(\theta)=P(\theta|M)$ is the prior. The evidence (marginal likelihood) $\mathcal{Z}$ is given by 

\begin{equation}\label{eq:evidence}
    \mathcal{Z}=P(D|M) = \int_{\Omega_\theta} \mathcal{L}(\theta)\pi(\theta)\mathrm{d}\theta,
\end{equation}
with the integral taken over the whole domain $\Omega_\theta$ of $\theta$. For a given set of data $D$, we can quantify how it supports one model $M_A$ compared to another $M_B$ by the ratio of evidences. This is also known as the \bfa:

\begin{equation}
    \label{eq:bayes}
    \mathcal{B}^A_B= \frac{P(D|M_A)}{P(D|M_B)}= \frac{\mathcal{Z}_A}{\mathcal{Z}_B}.
\end{equation}
Using Bayes' rule, we can relate the \bfa{} to the posterior ratio:

\begin{equation}
    \label{eq:posterior_ratio}
    \frac{P(M_A|D)}{P(M_B|D)}=\frac{P(D|M_A)}{P(D|M_B)}\frac{P(M_A)}{P(M_B)}=\mathcal{B}_B^A\frac{P(M_A)}{P(M_B)},
\end{equation}
where $P(M_i)$, $i = A,B$ is the prior probability of each model. \Cref{eq:posterior_ratio} shows that the posterior ratio is equivalent to the prior ratio `updated' by the \bfa{}, because the \bfa{} is the term that contains all information regarding the new data. In other words, the \bfa{} quantifies the `distinguishability power' of a set of data. A value of $\mathcal{B}^A_B > 1$ means that $M_A$ is more strongly supported by the data under consideration than $M_B$. \cite{kass1995} provide a guide for interpretation of \bfs{} and judge that $\mathcal{B}^A_B > 100$ (or $\log_{10}(\mathcal{B}^A_B) > 2$) can be regarded as being ``decisive''. We will assume this number as a threshold for confident model selection.

\subsection{Fiducial light curves}\label{sec:lightcurves}
There is no consensus in the early evolution of light curves of \ac{BNS} mergers. While one \ac{EM} counterpart of a \ac{BNS} merger was recorded with \gfo{}, this data set lacks \ac{UV} data before 15 hours \citep[][]{Evans2017} and optical data before 10 hours \citep[][]{Drout2017}. Even if earlier data was available, \ac{UVOIR} radiation from \ac{BNS} mergers more generally is also expected to be diverse depending on various binary and post-merger properties. We discuss this diversity and how it affects the conclusions of this study in more detail in \Cref{sec:discussion}. Being limited to \data{} of \gfo{}, the simulated light curves for the Bayesian study are based on \gfo{}.

\begin{figure}[htb!]
    \centering
    \includegraphics[width=0.5\textwidth]{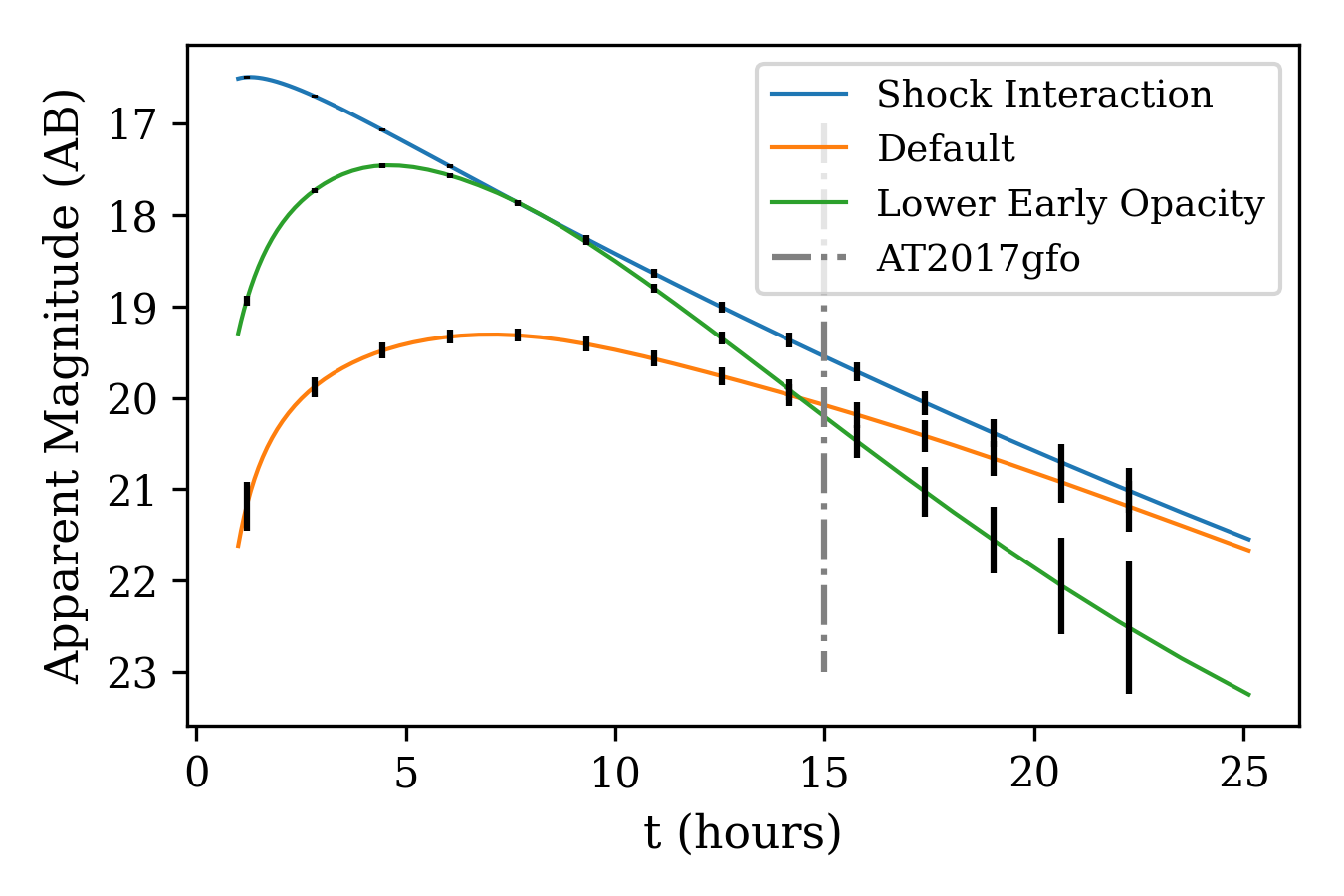}
    \caption{Comparison of \gfo-like light curves in the \D{} UV band (\SIrange[]{185}{215}{nm}) for \knmodels{} and \shmodel{} at luminosity distance $d_L = \SI{40}{Mpc}$. The data points correspond to 10 minute exposures that occur with a cadence of 97 minutes. There are fourteen data points in total for all light curves that are observed with identical timing. For reference, the dashdotted line indicates the timing of the first \ac{UV} data recorded of \gfo{} by \textit{Swift}/UVOT at 15 hours \citep[][]{Evans2017}.}
    \label{fig:comparison}
\end{figure}

We employ both radiation models discussed in Section \ref{sec:radiationmodels} to simulate the \data{}. Additionally, we use the \knmodel{} twice with different $v_{\rm max}$ and $\kappa_{\rm low}$, which allows for additional coverage of early \ac{UV} brightness allowed by \data{} obtained from \gfo. \Cref{fig:comparison} shows these three `fiducial light curves', and they are discussed in more detail in the remainder of this section.

The first fiducial light curve we call the \default{} light curve and is computed with the \knmodel{} using the same model parameters as  \cite{Hotokezaka2020} to fit \gfo{}. They found the resulting light curve to reasonably fit to the bolometric light curve and temperature data of \gfo{} \citep[taken from][]{Arcavi2018, Waxman2018}. The input parameters as well as prior distributions for the Bayesian analysis are given in \Cref{tab:parameters}.

\begin{deluxetable*}{llll}
    \tablecaption{\label{tab:parameters}Input parameters to the \kn{} and \sh{} models. On the left hand side, input parameter symbols are given along with units in parentheses. The prior density of the parameters are given in the third column, where the U(min, max) notation is used to indicate the minimum and maximum values of a uniform probability distribution. In the right hand column, the fiducial values for the fiducial light curves are given.}
    \tablecolumns{4}
    \tablewidth{0pt}
    \tablehead{
        \colhead{\textbf{Parameter (Unit)}} &
        \colhead{\textbf{Description}} &
        \colhead{\textbf{Prior Density}} &
        \colhead{\textbf{Fiducial value}}
    }
    \startdata
    \cutinhead{\textit{Default} [\textit{Lower Early Opacity}] \textit{Kilonova Model}}
    $M_{\mathrm{ej}}$ (M$_\odot$)     & Ejecta mass                                           & U(0.01, 0.1)                          & 0.05                    \\
    $v_{\mathrm{min}}$ (c)            & Minimum ejecta velocity                               & U(0.05, 0.2)                          & 0.1                     \\
    $v_{\mathrm{max}}$ (c)            & Maximum ejecta velocity                               & U(0.3, 0.8) [U(0.21, 0.8)]            & 0.4 [0.23]              \\
    $n_{\mathrm{ej}}$                & Power law index of ejecta density distribution        & U(3.5, 5)                             & 4.5                     \\
    $v_{\mathrm{transition}}$ (c)         & Transition velocity between high and low $\kappa$     & U($v_\mathrm{min}$, $v_\mathrm{max}$) & 0.2 [0.2]               \\
    $\kappa_\mathrm{high}$ (cm$^2$/g)  & Effective grey opacity for $v \leq v_\mathrm{\kappa}$ & U(1, 10)                              & 3                       \\
    $\kappa_\mathrm{low}$ (cm$^2$/g) & Effective grey opacity for $v \geq v_\mathrm{\kappa}$ & U(0.1, 1) [U(0.01, 0.1)]              & 0.5 [0.04]              \\
    \cutinhead{\textit{Shock Interaction Powered Model}}
    $M_{\mathrm{sh}}$ (M$_\odot$)     & Shocked ejecta mass                                   & U(0.005, 0.05)                        & 0.01                    \\
    $v_{\mathrm{sh}}$[c]             & Shocked ejecta minimum velocity                               & U(0.1, 0.3)                           & 0.2                     \\
    $R_0$ (10$^{10}$ cm)              & Initial shock radius                                  & U(1, 10)                              & 5                       \\
    $\kappa_\mathrm{sh}$ (cm$^2$/g)   & Effective grey opacity of shocked ejecta              & U(0.1, 1)                             & 0.5                     \\
    \enddata
\end{deluxetable*}

The second fiducial light curve is called the \bright{} light curve. Its inclusion here is motivated by recent work by \cite{Banerjee2020}  who found light curves with early UV brightness that peak 1.5--2 magnitude brighter than the \default{} model. Rather than using a (multiple zone) gray opacity model, they performed a radiative transfer simulation using opacities calculated by taking into account atomic structures, including highly ionized light r-process elements ($Z = 20 - 56$). The \bright{} light curve is a least squares fit with our \knmodel{} to both their simulated peak brightness and real \textit{Swift}/UVOT data from \gfo{} in AB magnitude. For this fit, we only allowed $v_{\rm transition}$, $v_{\rm max}$, $\kappa_{\rm low}$ to vary, which minimizes the change on the late time light curve while allowing for a good fit with the early \ac{UV} light curve of \cite{Banerjee2020}. The resulting $v_{\rm transition}$, $v_{\rm max}$, $\kappa_{\rm low}$ are shown in \Cref{tab:parameters}, but we find that $v_{\rm transition}$ remains the same. We also note that the AB magnitude of the \bright{} light curve at 0.1 day is brighter by $\sim$\,1 mag than the `light r-process+Sm+Nd+Eu' model  of \cite{Banerjee2022}, where they perform a radiative transfer simulation accounting for the expansion opacities of highly ionized lanthanide elements. Thirdly and lastly, a fiducial light curve is produced by the \shmodel{}. We use the fit to \gfo{} provided by \cite{Piro2018} assuming $s=3$. These parameters too are given in \Cref{tab:parameters}.

\subsection{Simulation of \data}\label{sec:data}
Using all the fiducial light curves, we simulate \data{} as observed by \textit{Dorado} in \ac{UV} and by \ac{LCO} in the optical r-band. While more optical bands (such as u, g, and i) are available, we found that their inclusion minimally affected model selection and parameter estimation (compared to just the r-band) and omitted them in exchange for reduced computational time.

\textit{Dorado} was a mission concept \citep{Singer2021} consisting of a SmallSat (slightly larger than a 12U CubeSat) spacecraft equipped with a 13 cm refractive (7 element) telescope with a 50 deg$^2$ \ac{FoV}. The instrument adopts a single (fixed) band-pass over the wavelength range from \SIrange[]{185}{215}{nm}. The \textit{Dorado} camera employs delta-doped \ac{CCD} detectors to provide surface passivation and reflection-limited response over the \ac{UV} bandpass \citep[][]{Nikzad2017}. The spacecraft was designed to accommodate a wide range of low-Earth orbits and given ride-share availability we assume a noon-midnight sun synchronous (polar) orbit with a 600 km altitude for the simulations described here.

For the simulation of \textit{Dorado} \data, representative observational cadences were derived using the \textit{dorado-scheduling}\footnote{\url{https://github.com/nasa/dorado-scheduling/}} software package. \textit{dorado-scheduling} can generate realistic observational sequences for gravitational-wave localizations, including exclusion constraints (Sun, Moon, and Earth limb), as well as satellite down time for passages through the South Atlantic Anomaly. The software can define optimized observing strategies based on the GW localization region and distance, as well as a (position-dependent) exposure time calculator (see below). However, for these simulations we assumed the first data point is observed at $\Delta t = 72$\,min post merger (to allow for time to uplink to the spacecraft), and a fixed exposure time of 10 minutes every orbit (97 minute cadence) on a single pointing.

Signal-to-noise estimates were derived using the \textit{dorado-sensitivity}\footnote{\url{https://github.com/nasa/dorado-sensitivity/}}, the exposure time calculator for the \textit{Dorado} mission. \textit{dorado-sensitivity} generates realistic foreground models for zodiacal light (based on spacecraft and target location), as well as airglow emission (based on location within the orbit). While the software is capable of incorporating Milky Way dust extinction, this is not included here. The fiducial light curve is folded through the \textit{Dorado} effective area curve to generate the expected source counts, and then compared with the background (including both foreground and source shot noise). For reference, for a 10 minute exposure, \textit{Dorado}’s 5$\sigma$ limiting magnitude for an isolated point source is typically $\sim$\,20.5\,mag (AB).

For \gbo{} observations, we similarly use the same detection schedule for all simulated events but starting at 12 hours and with a 12 hour cadence up to 48 hours. To calculate this \data{} we use an \ac{ETC} constructed using data of \ac{LCO} instruments\footnote{\url{https://exposure-time-calculator.lco.global/}}. \ac{LCO} is a network of 25 telescopes at seven sites around the world, sensitive in optical and \ac{NIR} wavelengths. Being purpose-built to observe transient events, it robotically schedules observations and leverages its global network to observe around the clock and make rapid observations of \acp{ToO}, evading potential local weather limitations \citep[][]{Brown2013}. 

\section{Bayesian Analysis}\label{sec:results}
This study employs \textit{Dynesty} \citep[][]{Speagle2020} for Nested Sampling \citep[][]{Skilling2004,Skilling2006}. Nested Sampling is a method for simultaneously estimating posterior probability $P(\theta|D,M)$ (see Equation  \ref{eq:posterior}) and evidence $\mathcal{Z}$ (see Equation \ref{eq:evidence}). The full pipeline for simulation of \data{} and subsequent Bayesian analysis is publicly available on Github\footnote{\url{https://github.com/Basdorsman/kilonova-bayesian-analysis/}}. This section presents the results of the Bayesian analysis, and is structured as follows: \Cref{sec:UV} presents the results of model selection and parameter estimation via \data{} consisting of satellite-based \ac{UV}, ground-based optical, and both. \Cref{sec:delays} considers satellite-based \ac{UV} \data{} but analyzes the effect of delayed observation of the target.

\subsection{Model Selection and Parameter Estimation for UV, Optical and Joint Data}\label{sec:UV}

\begin{figure*}[htb!]
    \centering
    \includegraphics[width=0.8\textwidth]{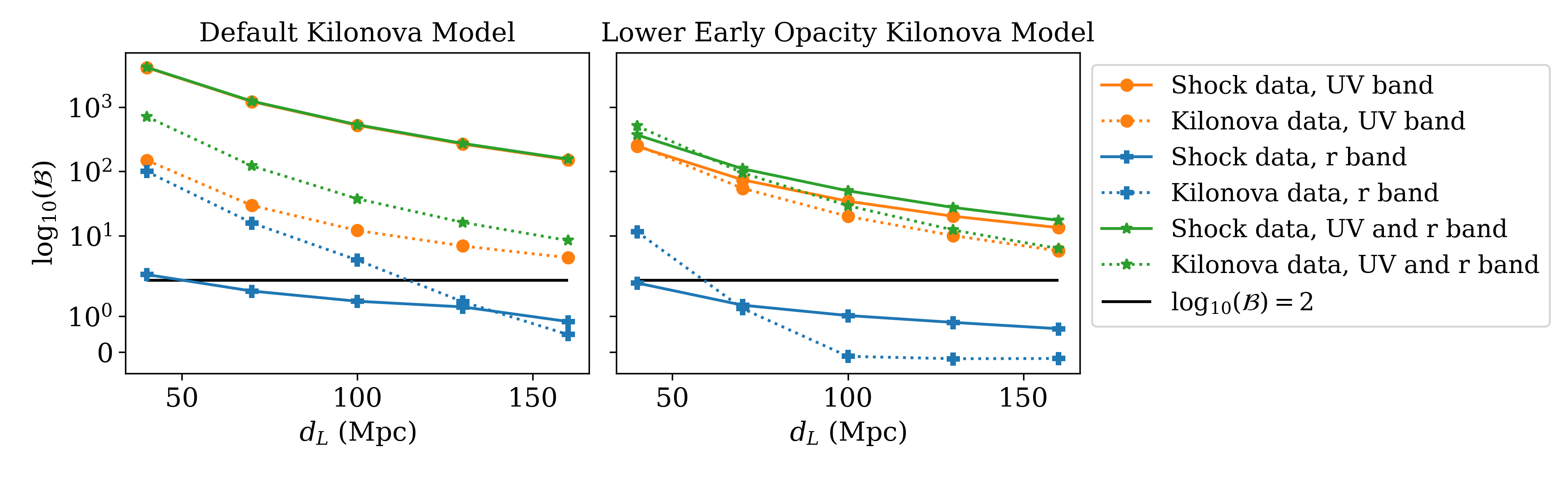}
    \caption{\bfs{} obtained from employing \sbu, \gbo{} or joint \ac{UVO} \data{}. \textit{Left}: Resulting \bfs{} when attempting to distinguish the \shmodel{} from \default{} \knmodel{}. Note here that the results obtained from shock data in \ac{UV}+r bands overlap the results from the \ac{UV} band. \textit{Right}: Same, but now distinguishing the \shmodel{} from the \bright{} \knmodel{}. The solid (dotted) lines correspond to \bfs{} resulting when using simulated data produced by the \shmodel{} (\knmodel). The black horizontal line indicates the  $\log_{10}(\mathcal{B})=2$ threshold for decisive evidence.}
    \label{fig:bayes_bands}
\end{figure*}

\Cref{fig:bayes_bands} shows \bfs{} as a function of luminosity distance of the merger. In all cases up to \SI{160}{Mpc}, the \knmodels{} can be confidently distinguished from the \shmodel{} using only \ac{UV} data. Conversely, the optical data is able to distinguish the models only up to $\sim$\,110 (60) \si{Mpc} for the \default{} (\bright{}) light curves. Moreover, optical data produced by the \sh{} light curve presents edge cases at \SI{40}{Mpc} and beyond this distance the data is insufficient to distinguish the models. The combined \ac{UVO} data set allows for (marginally) better distinguishability than the \ac{UV} data.

\begin{figure*}[htb!]
    \centering
    \includegraphics[width=0.7\textwidth]{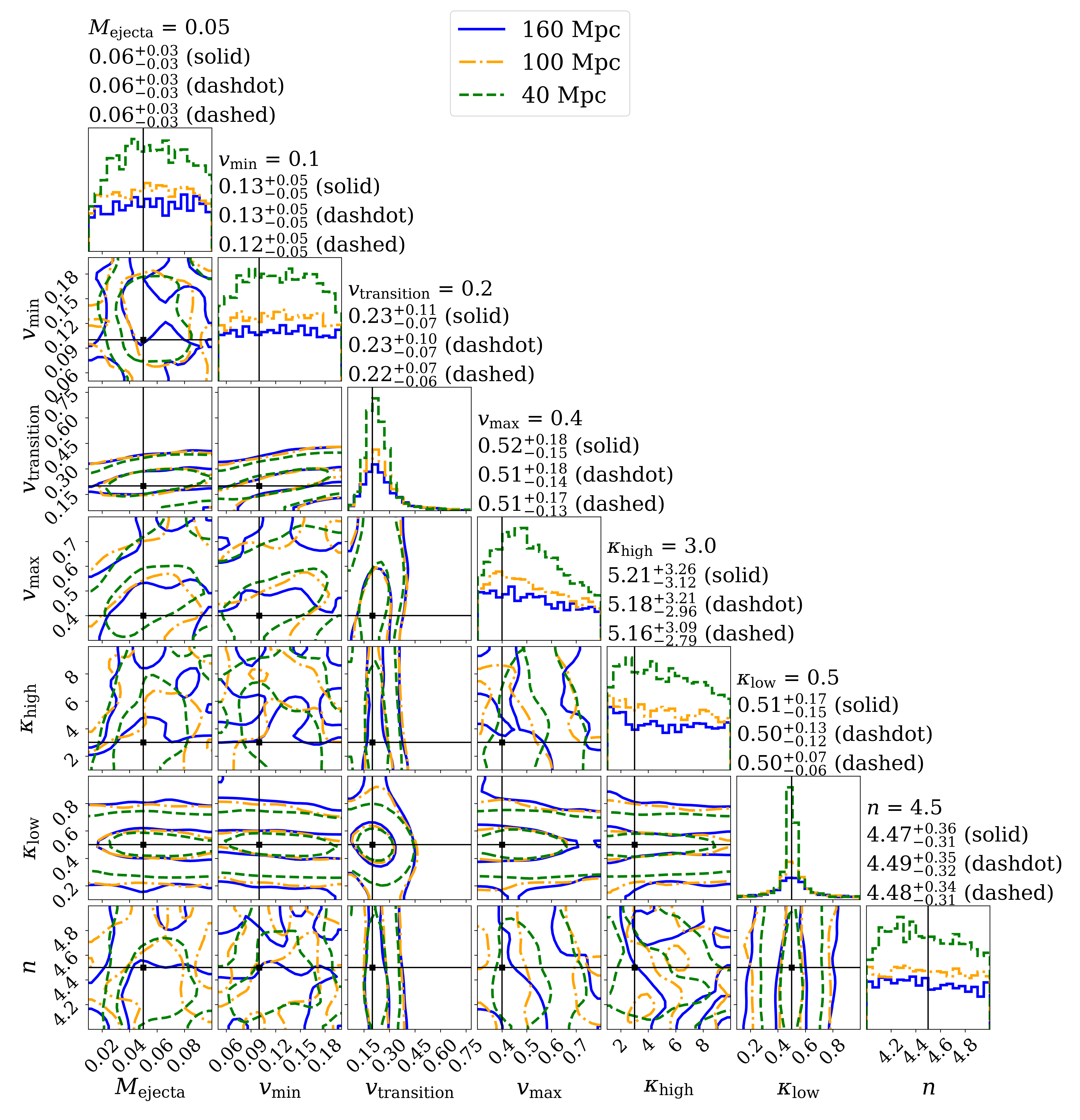}
    \caption{Posterior distributions of the parameters of the \default{} \knmodel, for an \gfo-like event at \distances. The sub-figures on the diagonal of the corner plots display the \acp{PPD} and the fiducial values for the model parameters (black lines). The sub-figure titles on the diagonal list the 0.16, 0.5, and 0.84 fractional quantiles. The contours in the \ac{2D} plots indicate the \ac{2D} 1$\sigma$ (39\%) and 2$\sigma$ (86\%) credible regions.}
    \label{fig:cornerkilonovadefault}
\end{figure*}

To get a handle on parameter estimation and its distance dependency, \Cref{fig:cornerkilonovadefault} shows the posterior probability distributions (\acp{PPD}) for the \default{} \knmodel{} at three selected luminosity distances: \distances. Two out of the seven model parameters: $v_\mathrm{transition}$ and $\kappa_\mathrm{low}$, are well constrained by the \ac{UV} \data{}, but especially so at the most close-by distance at \SI{40}{Mpc}. We note that these two parameters in part define the outer opacity zone of the ejecta outflow. Because the outer opacity zone is bright in \ac{UV} (for these parameter values such that the light curves resemble \gfo{}) we expect these parameters to be relatively well constrained by the \ac{UV} \data{}. The remaining 5 parameters are not significantly constrained.

\begin{figure}[htb!]
    \centering
    \includegraphics[width=0.5\textwidth]{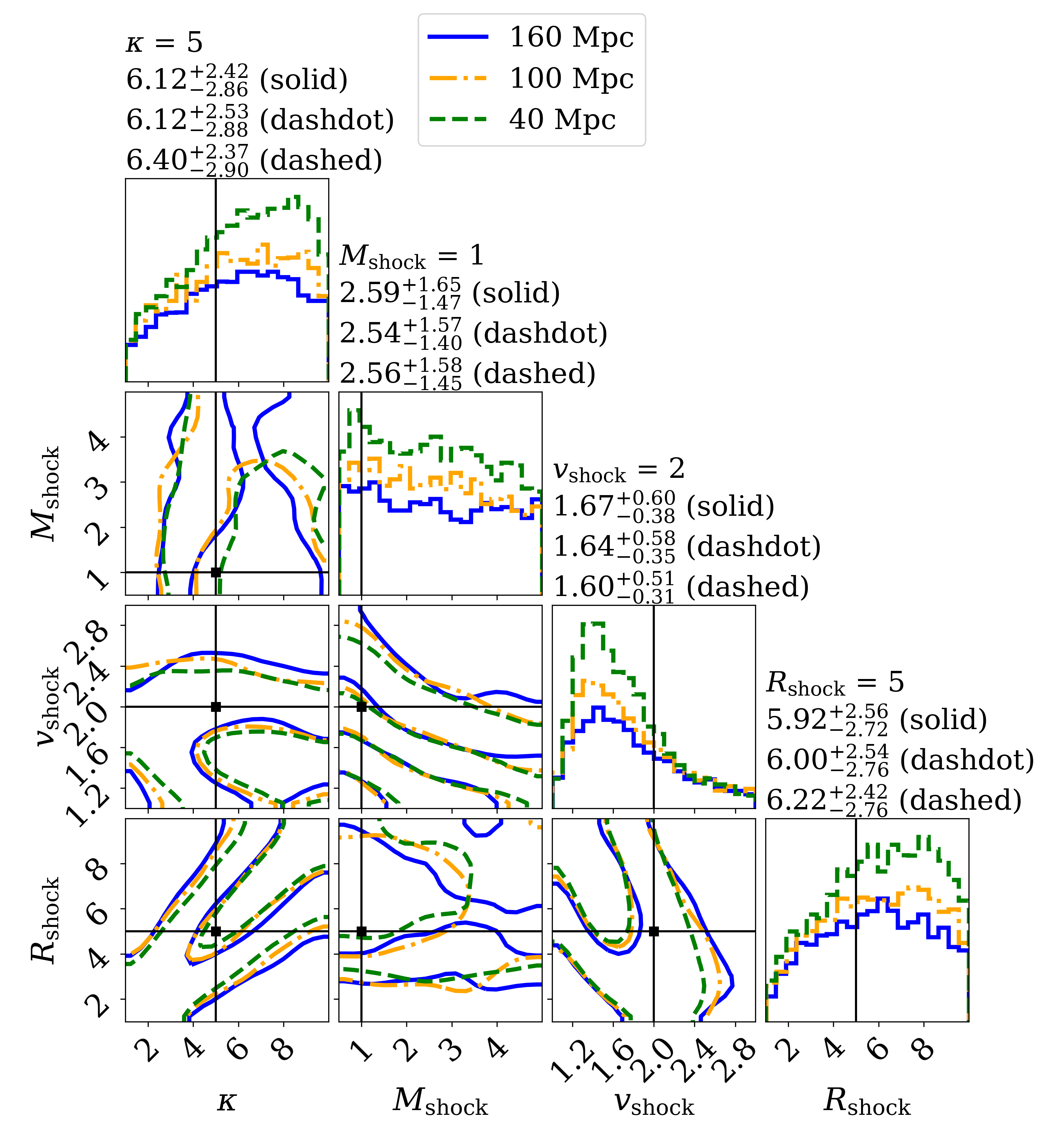}
    \caption{Same as \Cref{fig:cornerkilonovadefault} but for the \shmodel.}
    \label{fig:cornershock}
\end{figure}

\Cref{fig:cornershock} shows the \ac{PPD} for the \shmodel. Even for \SI{40}{Mpc} the parameters are constrained only slightly within prior ranges. As is evident from the figure, this lack in constraining of parameters is due to parameter degeneracy in this model. We find the following degeneracy in the model: consider a set of parameters $M_1$, $\kappa_1$ and $R_1$ that produce some $L_1$, $T_{\mathrm{eff},1}$ and $r_{\mathrm{ph},1}$. There exists another set of parameters, $M_2 = \phi^{-1}M_1$, $\kappa_2 = \phi \kappa_1$ and $R_2 = \phi R_1$, where $\phi$ is some constant. Substituting these into \Cref{eq:Lshock,eq:Tshock,eq:rshock}, all factors $\phi$ cancel and we are left with $L_1 = L_2$, $T_{\mathrm{eff},1}=T_{\mathrm{eff},2}$ and $r_{\mathrm{ph},1}=r_{\mathrm{ph},2}$. This means that the blackbody spectrum is identical for different parameters. Because of this degeneracy, parameters cannot be individually constrained even if observing in multiple bands. We note however, that in the case of \gfo{}, the $R_\mathrm{shock}$ could be independently constrained from the time delay between \acp{GW} and \ac{GRB}. In such a case $\kappa$ would become constrained, but a degeneracy remains between $M_{\rm shock}$ and $v_{\rm shock}$ ($M_2 = \phi^{-4}M_1$ and $v_2 = \phi v_1$).

\begin{figure*}[htb!]
    \centering
    \includegraphics[width=0.7\textwidth]{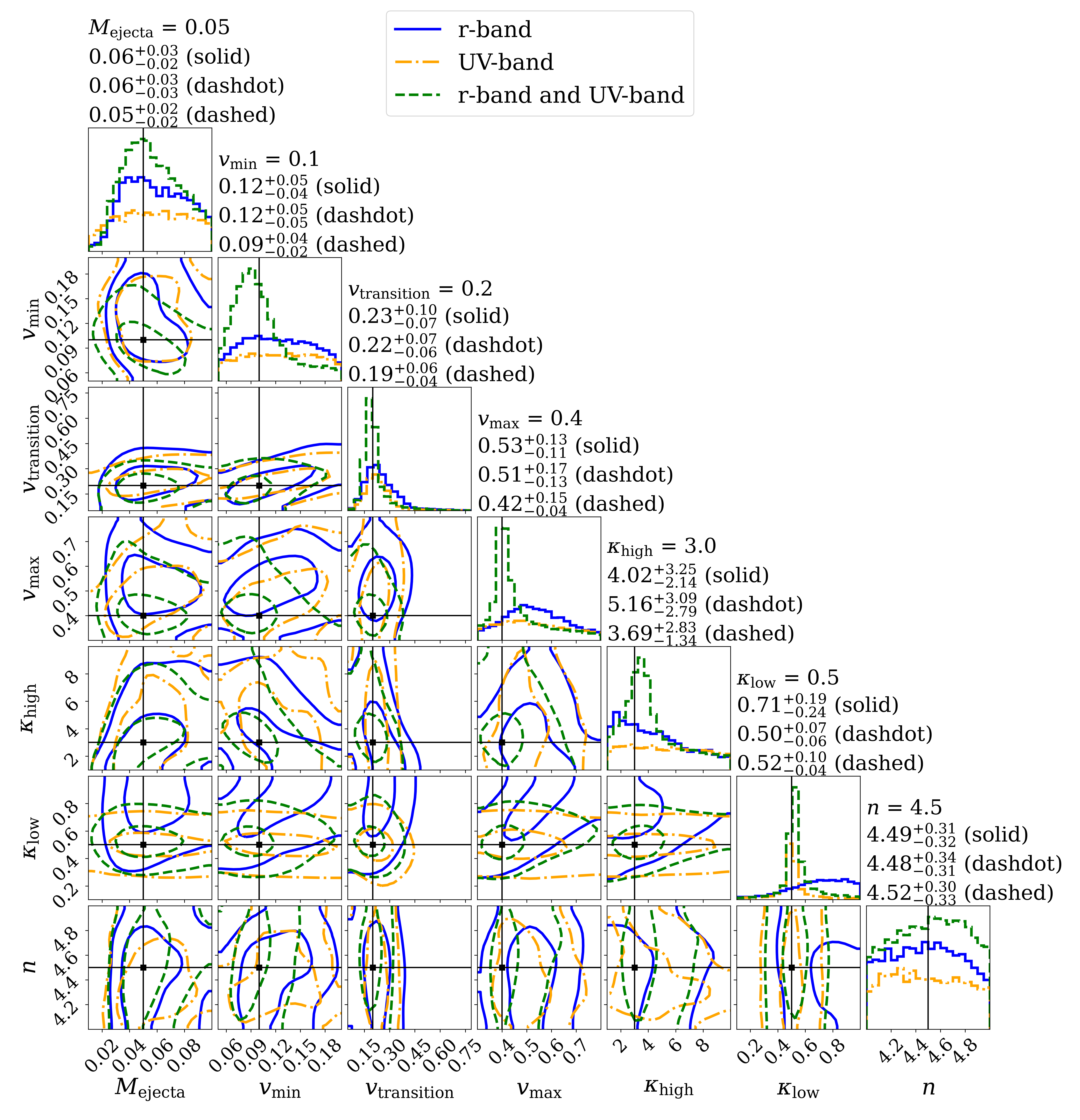}
    \caption{Similar to \Cref{fig:cornerkilonovadefault}, but now comparing results for parameter estimation using \gbo{} \data{} (solid), using \sbu{} \data{} (dashdot) and both (dashed) for an event at \SI{40}{Mpc}.}
    \label{fig:variousbands}
\end{figure*}

\Cref{fig:variousbands} shows the results for the \default{} \knmodel{} again (same as \SI{40}{Mpc} in \Cref{fig:cornerkilonovadefault}), but now also including a comparison with \gbo{} \data{} and with the combined data of both bands. In isolation ground-based optical observations constrain $M_\mathrm{ejecta}$ and $\kappa_\mathrm{high}$ more accurately and precisely than the \ac{UV}. For $v_\mathrm{min}, v_\mathrm{transition}, v_\mathrm{max}, \kappa_\mathrm{high}$ and $n$, performance of either single band is comparable, but constraints on $\kappa_\mathrm{low}$ are much worse for the optical than \ac{UV}. However, in combination ground-based optical and \ac{UV} are complementary to each other, providing significant improvement compared to using either band in isolation. In that case, all parameters except $n$ are well constrained.

\subsection{Delayed detection}\label{sec:delays}
\begin{figure*}[htb!]
    \centering
    \includegraphics[width=0.8\textwidth]{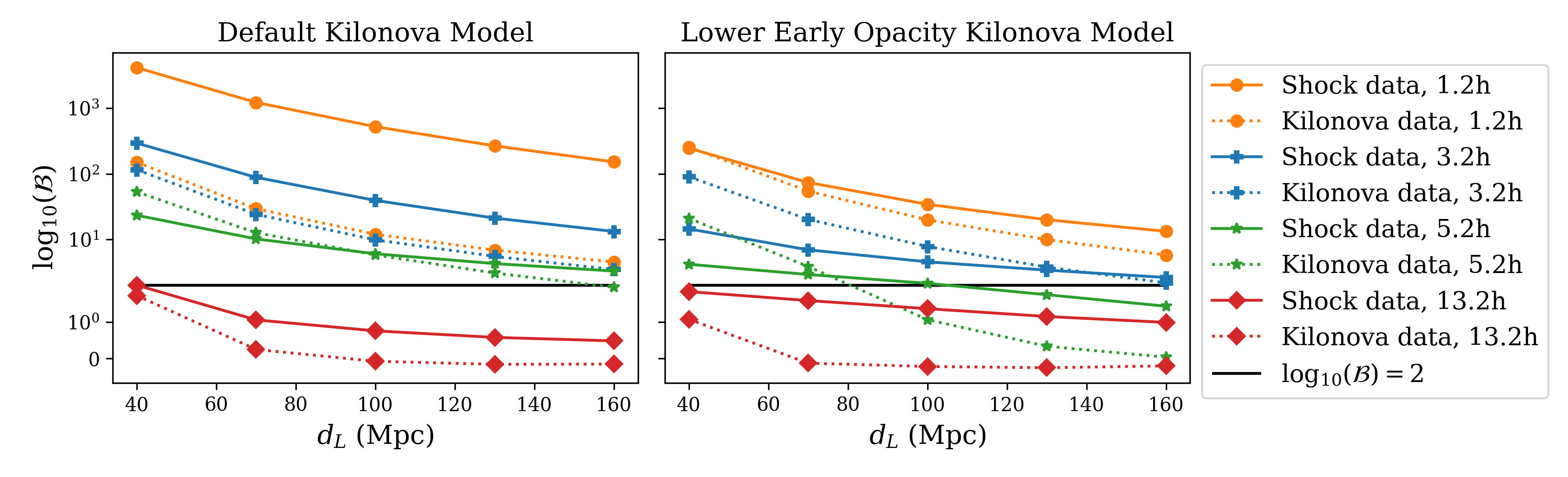}
    \caption{\bfs{} obtained from model selection using \sbu{} \data. The data is shifted in time such that the first data point is obtained at \num{1.2}, \num{3.2}, \num{5.2} and \num{13.2} hours after \ac{GW} trigger. The distinction between left and right, dotted and solid lines as well as the black horizontal line are explained in the caption of \Cref{fig:bayes_bands}.}
    \label{fig:bayesdelays}
\end{figure*}

The time it takes from \ac{GW} trigger to first on-target exposure (effective response time) should be as short as possible to capture the physics that govern the early post-merger system. This is especially relevant for the \ac{UV} compared to optical and \ac{IR} because the emission rises and fades more rapidly ($\sim$\,hours compared to $\sim$\,days and $\sim\,$weeks respectively). \Cref{fig:bayesdelays} shows the results for model selection using \sbu{} \data, where the data set as a whole was shifted forwards in time with various delays as indicated in the figure. Note that the cases for data at \SI{1.2}{h} are identical to the \ac{UV} data in \Cref{fig:bayes_bands}. For distances up to \SI{160}{Mpc}, it is possible to distinguish the models if the first data is collected up to 5.2 (3.2) hours after \ac{GW} trigger for the \default{} (\bright) light curves. For \SI{40}{Mpc} and beyond, the models cannot be distinguished if the first data is collected after 13.2 hours.

\begin{figure*}[htb!]
    \centering
    \includegraphics[width=0.7\textwidth]{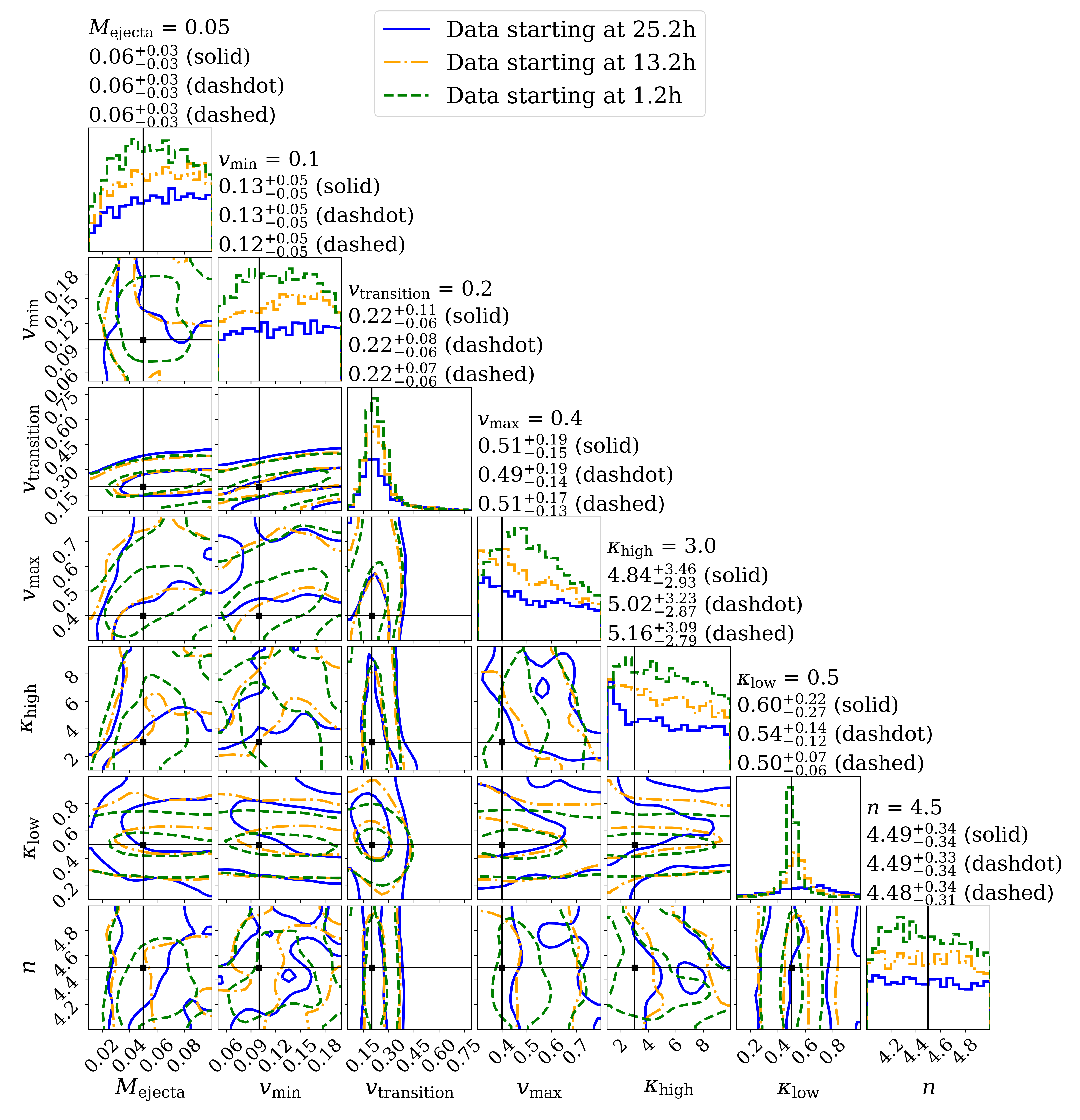}
    \caption{Similar to \Cref{fig:cornerkilonovadefault}, but now with the \sbu{} \data{} at \SI{40}{Mpc} shifted in time such that the first data point is obtained at \num{1.2}, \num{13.2} and \num{25.2} hours.}
    \label{fig:cornerdelays}
\end{figure*}

\Cref{fig:cornerdelays} shows the results for parameter estimation for increasingly delayed \data. Note that the results for \SI{1.2}{h} are identical to \SI{40}{Mpc} in \Cref{fig:cornerkilonovadefault}. All parameters experience worsening constraints for increasing response time. Of the two parameters that were previously well constrained for \sbu{} \data, $v_\mathrm{transition}$ is still well constrained even for data starting at \SI{25.2}{h} while $\kappa_\mathrm{low}$ is not well constrained beyond \SI{13.2}{h}. We posit that $v_\mathrm{transition}$ may be well constrained even with such delayed data because it affects both opacity zones, while $\kappa_\mathrm{low}$ only plays a role in defining the quicker fading outer opacity zone.

\section{Discussion and Conclusion}\label{sec:discussion}
We have performed model selection and parameter estimation, assuming \gfo-like `blue' emission produced by either \kn{} or \sh{} radiation and assuming observational capabilities of a \ac{UV} satellite (\textit{Dorado}) and a ground-based observation network (\ac{LCO}). Our main findings are threefold: 

Firstly, our results suggest that \sbu{} \data{}, unaided by \gbo{} follow-up would be sufficient to distinguish these models for an event up to at least \SI{160}{Mpc} for all considered scenarios. Conversely, \gbo{} \data{} only selects models up to $\sim$\,\SI{110}{Mpc} in the most optimistic scenario and has considerably worse performance ($\sim$\,\SI{40}{Mpc}) in other scenarios. Combined \ac{UVO} \data{} only marginally improves model selection beyond what is possible with only \ac{UV} data.

Secondly, we find that combined \ac{UVO} \data{} allows the full set of parameters (save $n$, which defines steepness of the ejecta velocity profile) of our \knmodel{} to be well constrained. In comparison, \data{} from either \ac{UV} or optical alone constrains smaller subsets of the parameters and with reduced accuracy and precision. This result suggests that multiple-wavelength detections are essential for constraining early ejecta geometry and opacity.

Thirdly, we find that the data allows to discern between the models if first data is observed no later than 3.2 (5.2) hours after \ac{GW} trigger for the \default{} (\bright{}) light curves up to \SI{160}{Mpc}. Moreover, after \SI{13.2}{h} the models can no longer be discerned beyond \SI{40}{Mpc}. These results indicate that rapid on-target observations on the order of a few hours is necessary for distinguishing the \kn{} from \sh{} models through \ac{UV} \data{}.

To place these results in broader context, we comment on prospects for planned wide-field \ac{UV} instruments to detect and characterize \ac{EM} counterparts during the O5 observation run of \ac{GW} observatories \ac{HLVK} in 2024 and beyond. We focus here on O5, because the launch dates of ULTRASAT (2025) and UVEX ($\sim$\,2028) are in part overlapping with O5.

Simulations of \ac{GW} detections in O5 were done by \cite{Petrov2022}. From their simulated data set \citep{leo_singer_2021_4765752} we compute that of \ac{BNS} events within 160 Mpc, 69\% are localized within 100 deg$^2$, which would be easily followed up by \textit{Dorado} within a single orbit. The fraction of events that meet both distance and localization criteria, multiplied by the annual detection rate (190, from \cite{Petrov2022}) gives an annual detection rate of $\sim$\,3.2. If we include events up to 400 Mpc, we find that still 46\% of events are localized within 100 deg$^2$, corresponding to an annual rate of $\sim$\,20 events. Thus, for \ac{BNS} mergers, if we assume similar brightness to \gfo{} (see below), we expect an annual detection rate of $\sim$\,$3.2$ of which \sh{} and \kn{} radiation could be distinguished and \kn{} ejecta parameters constrained with a \textit{Dorado}-like satellite. For \ac{BHNS} mergers, which have not been simulated here, the expected number of detected EM counterparts should be much lower, as only a small region of BHNS parameter space will lead to disruption of the NS \citep[][]{Foucart2020}. While the simulations performed here are specific to \textit{Dorado}, some of the findings also are relevant to currently planned wide-field \ac{UV} missions. To start with, the wide-field \ac{UV} mission ULTRASAT is more sensitive than \textit{Dorado} with a limiting magnitude of $\sim$\,22.3 (5$\sigma$, AB, 3$\times$\SI{300}{s} exposure) and has a larger \ac{FoV} at 200 deg$^2$ \citep[][]{Asif2021}. Because of this, and ULTRASATs ability to point to a given \ac{ToO} within 30 minutes \citep[][]{Sagiv2014}, it is both sensitive and quick enough to provide the \ac{UV} data for model selection and constraints as suggested by simulations here. Secondly, UVEX is even more sensitive with a limiting magnitude of 25 (AB, 5$\sigma$) but has a smaller \ac{FoV} of 12 deg$^2$. In addition, UVEX will provide additional detail in the light curves and spectra through its two (far and near UV) sensors and on-board spectroscope \citep[][]{Kulkarni2021}. Because this mission has additional capabilities, applicability of these results to this mission is not straightforward. For both we recommend separate simulations that accurately represent their mission parameters to assess capability for model distinction and parameter estimation.

Finally, we remark on the various simplifying assumptions made in this study that may affect the robustness of above findings. To start with, throughout this work we have made the assumption that the \ac{EM} counterparts of \ac{BNS} mergers all are similar to that of \gfo. This introduces a `\gfo-bias' in our predictions for the ability of a future UV satellite to achieve science goals. It is as of yet uncertain how representative \gfo{} is for the actual kilonova population and O4 and O5 are expected to shed light on this. This population is expected to be diverse, for example due to binary properties such as mass ratio and spins, but also post-merger properties such as remnant outcome \citep[e.g.][]{Kawaguchi2020} and jet-ejecta interaction \citep[e.g.][]{Klion2021}. These are expected to have an effect on ejecta geometry and composition which in turn affects brightness and color of light curves.

Another assumption underlying this study is that both models assume isotropic ejecta. However, the photon emission as well as composition and geometry of the ejecta are not expected to be spherically symmetric \citep[e.g.][]{Metzger2017, Heinzel2021}. Other models that include the inclination angle, such as the grid of 2D simulations presented by \cite{Wollaeger2021}, would allow for a more representative study covering the angular dependence of \ac{UV} light curves. Still, \cite{Heinzel2021} recommend the inclusion of $\sim$\,1 mag uncertainties for \kn{} models used in Bayesian studies relating to inclination angle to capture yet unknown systematic model uncertainties. 

We also note that a significant source of uncertainty remains in the nuclear physics taking place in these high energy events. For example, \cite{Zhu2022} find that uncertainties in nuclear inputs lead to typically one order of magnitude variation in inferred nuclear heating, bolometric luminosity and ejecta mass. 

Lastly, although the scope of this study was limited to \sh{} and \kn{} radiation, we note that a more comprehensive study may include additional radiation scenarios, such as a neutron precursor or winds driven by a long-lived remnant, and also a combination of emission channels.

In conclusion, these results show that \ac{UV} data offers a unique window to distinguish the processes governing the early post-merger system. For \textit{Dorado}, rapid follow-up as well as the ability to quickly localize the target within a few hours catches the quickly fading \ac{UV} emission, allowing to distinguish models. We also find that through multi-wavelength observations the kilonova emission can be constrained up to at least \SI{160}{Mpc}, unlocking a fuller understanding of the geometry and opacity of the ejecta outflow.\\

BD is grateful to the whole \textit{Dorado} team for their development of the mission concept, which enabled this detailed case study. BD acknowledges support from ERC Consolidator grant No. 865768 AEONS (PI: Anna Watts).

\software{Python~ language~ \citep{Oliphant2007}, NumPy~\citep{vanderWalt2011}, SciPy~\citep{Jones2001}, Matplotlib~\citep{Hunter2007,Droettboom2018}, IPython~\citep{Perez2007}, Jupyter~\citep{Kluyver2016}, Dynesty~\citep{Speagle2020}, Corner.py~\citep{Foreman-Mackey2016}, GNU Parallel~\citep{Tange2021}}

\bibliography{bibliography}{}
\bibliographystyle{aasjournal}

\end{document}